\documentclass[aps,prb,preprint,nopacs,superscriptaddress,]{revtex4}

\usepackage{graphicx}
\usepackage{verbatim}
\usepackage{mathrsfs}
\usepackage{gensymb}
\usepackage{color}
\usepackage{ulem}
\usepackage{times}

\usepackage{etoolbox}
\pretocmd{\abstractname}{\newpage}{}{}

\usepackage{amsmath,amsfonts,amssymb}
\usepackage{graphicx}

\newcommand{\bee}{\begin{equation}}
\newcommand{\ee}{\end{equation}}
\newcommand{\ef}{$E_\textrm{F}$}

\newcommand{\pana}{a}
\newcommand{\panb}{b}
\newcommand{\panc}{c}
\newcommand{\pand}{d}
\newcommand{\pane}{e}
\newcommand{\panf}{f}
\newcommand{\pang}{g}

\def\3{2.5in}    
\def\2{2.5in}
\def\4{3.0in}\def \beq {\begin{equation}}
\def \eeq {\end{equation}}
\begin{document}

\title{Discovery of topological chiral crystals with helicoid arc states}
\author{Daniel S. Sanchez}
\affiliation {Laboratory for Topological Quantum Matter and Spectroscopy (B7), Department of Physics, Princeton University, Princeton, New Jersey 08544, USA}

\author{Ilya Belopolski}
\affiliation {Laboratory for Topological Quantum Matter and Spectroscopy (B7), Department of Physics, Princeton University, Princeton, New Jersey 08544, USA}

\author{Tyler A. Cochran}
\affiliation {Laboratory for Topological Quantum Matter and Spectroscopy (B7), Department of Physics, Princeton University, Princeton, New Jersey 08544, USA}

\author{Xitong Xu}
\affiliation{International Center for Quantum Materials, School of Physics, Peking University, China}

\author{Jia-Xin Yin}
\affiliation {Laboratory for Topological Quantum Matter and Spectroscopy (B7), Department of Physics, Princeton University, Princeton, New Jersey 08544, USA}

\author{Guoqing Chang}
\affiliation {Laboratory for Topological Quantum Matter and Spectroscopy (B7), Department of Physics, Princeton University, Princeton, New Jersey 08544, USA}

\author{Weiwei Xie}
\affiliation{Department of Chemistry, Louisiana State University, Baton Rouge, LA, 70803, USA}

\author{Kaustuv Manna}
\affiliation{Max Planck Institute for Chemical Physics of Solids, Dresden, D-01187, Germany}

\author{Vicky S\"{u}{\ss}}
\affiliation{Max Planck Institute for Chemical Physics of Solids, Dresden, D-01187, Germany}

\author{Cheng-Yi Huang}
\affiliation{Institute of Physics, Academia Sinica, Taipei 11529, Taiwan}

\author{Nasser Alidoust}
\affiliation {Laboratory for Topological Quantum Matter and Spectroscopy (B7), Department of Physics, Princeton University, Princeton, New Jersey 08544, USA}
\affiliation {Rigetti Computing, Berkeley, California 94720, USA}

\author{Daniel Multer}
\affiliation {Laboratory for Topological Quantum Matter and Spectroscopy (B7), Department of Physics, Princeton University, Princeton, New Jersey 08544, USA}
\affiliation {Department of Chemistry, Princeton University, Princeton, New Jersey 08544, USA}

\author{Songtian S. Zhang}
\affiliation {Laboratory for Topological Quantum Matter and Spectroscopy (B7), Department of Physics, Princeton University, Princeton, New Jersey 08544, USA}

\author{Nana Shumiya}
\affiliation {Laboratory for Topological Quantum Matter and Spectroscopy (B7), Department of Physics, Princeton University, Princeton, New Jersey 08544, USA}

\author{Xirui Wang}
\affiliation {International Center for Quantum Materials, School of Physics, Peking University, China}

\author{Guang-Qiang Wang}
\affiliation {International Center for Quantum Materials, School of Physics, Peking University, China}

\author{Tay-Rong Chang}
\affiliation{Department of Physics, National Cheng Kung University, Tainan 701, Taiwan}

\author{Claudia Felser}
\affiliation{Max Planck Institute for Chemical Physics of Solids, Dresden, D-01187, Germany}

\author{Su-Yang Xu}
\affiliation {Laboratory for Topological Quantum Matter and Spectroscopy (B7), Department of Physics, Princeton University, Princeton, New Jersey 08544, USA}

\author{Shuang Jia}
\affiliation{International Center for Quantum Materials, School of Physics, Peking University, China}\affiliation{Collaborative Innovation Center of Quantum Matter, Beijing,100871, China}\affiliation{CAS Center for Excellence in Topological Quantum Computation, University of Chinese Academy of Science, Beijing 100190, China}

\author{Hsin Lin}
\affiliation{Institute of Physics, Academia Sinica, Taipei 11529, Taiwan}


\author{M.~Zahid~Hasan$^{\dag}$\footnote[0]{$^{\dag}$Corresponding author (email): mzhasan@princeton.edu
\\\textbf{Submitted on August 10, 2018}; References updated. }}
\affiliation{Laboratory for Topological Quantum Matter and Spectroscopy (B7), Department of Physics, Princeton University, Princeton, New Jersey 08544, USA}
\affiliation{Lawrence Berkeley National Laboratory, Berkeley, California 94720, USA}
\maketitle

\textbf{The quantum behaviour of electrons in materials lays the foundation for modern electronic and information technology \cite{natnews, kmoore, revHK, revQZ, jia, rev4, rev6, rev6a, rev7, KramersWeyl,News}. Quantum materials with novel electronic and optical properties have been proposed as the next frontier, but much remains to be discovered to actualize the promise. Here we report the first observation of topological quantum properties of chiral crystals \cite{KramersWeyl,News} in the RhSi family. We demonsrate that this material hosts novel phase of matter exhibiting nearly ideal topological surface properties that emerge as a consequence of the crystals' structural chirality or handedness \cite{KramersWeyl,News}. We also demonstrate that the electrons on the surface of this crystal show a highly unusual helicoid structure that spirals around two high-symmetry momenta signalling its topological electronic chirality. Such helicoid Fermi arcs on the surface experimentally characterize the topological charges of $\pm{2}$, which arise from the bulk chiral fermions. The existence of bulk high-fold degenerate fermions are guaranteed by the crystal symmetries, however, in order to determine the topological charge in the chiral crystals it is essential to identify and study the helical arc states. Remarkably, these topological conductors we discovered exhibit helical Fermi arcs which are of length $\pi$, stretching across the entire Brillouin zone and orders of magnitude larger than those found in all known Weyl semimetals \cite{rev4, rev6, jia, rev6a}. Our results demonstrate novel electronic topological state of matter on a structurally chiral crystal featuring helicoid Fermi arc surface states. The exotic electronic chiral fermion state realised in these materials can be used to detect a quantised photogalvanic optical response or the chiral magnetic effect and its optical version in future devices as described by G. Chang \textit{et.al.,} ``Topological quantum properties of chiral crystals" Nature Mat. 17, 978-985 (2018) \cite{KramersWeyl}.}

The discovery of topological insulators has inspired the search for a wide variety of topological conductors \cite{natnews, kmoore, revHK, revQZ, rev4, rev6, jia, rev7, NielsenNinomiya1, filling_constraint1, rev6a, Topo.gapless.phase, Wan, HgCr2Se4, News, Weyl-Multilayer, unconventionalWeyl, RhSi, CoSi, TaAs1, TaAs2, ARPES-TaAs1, KramersWeyl, Ben1, Manes, ARPES-TaAs2, Nobel}. One example of topological conductor is the Weyl semimetal (WSM), featuring emergent Weyl fermions as low-energy excitations of electrons. These Weyl fermions are associated with topological chiral charges (Chern numbers) that locate at two-fold degenerate band crossings \cite{natnews, kmoore, rev4, rev6, rev6a, rev7, NielsenNinomiya1, Topo.gapless.phase, Wan, Weyl-Multilayer} in momentum space. In general, such emergent topological chiral fermions may appear in a variety of types including not only two-fold degenerate Weyl fermions \cite{rev4, rev7, rev6, rev6a, NielsenNinomiya1, Topo.gapless.phase, Wan,Weyl-Multilayer, HgCr2Se4}, Kramers-Weyl fermions \cite{KramersWeyl, News} or higher-fold fermions \cite{unconventionalWeyl, RhSi, CoSi}. Recently, a few non-centrosymmetric crystals were identified where a band inversion gives rise to a WSM state \cite{TaAs1, TaAs2, ARPES-TaAs1, ARPES-TaAs2, rev4, rev7, rev6}. However, all these materials suffer from several drawbacks: a large number of Weyl fermions, Weyl fermions close to each other in momentum space, and short Fermi arcs which are much (orders of magnitude) less topologically robust. In order to thoroughly explore and utilise the robust and unusual quantum phenomena induced by chiral fermions in optics or magneto-transport, topological conductors with near-ideal electronic properties or novel types of topological conductors are needed \cite{natnews, kmoore}.

A different approach toward searching for ideal topological conductors is to examine crystalline symmetries, which can also lead to topological band crossings \cite{revHK, revQZ, rev6}. For instance, it has been shown that non-symmorphic symmetries can guarantee the existence of band crossings for certain electron fillings \cite{unconventionalWeyl, filling_constraint1, Ben1}. As another example, we might consider structurally chiral crystals whose lattice possess no inversion, mirror and roto-inversion symmetries \cite{KramersWeyl}. Structurally chiral crystals are expected to host a variety of topological band crossings which are guaranteed to be pinned to time-reversal invariant momenta (TRIMs) \cite{KramersWeyl, Manes}. Moreover, structurally chiral topological crystals naturally give rise to a quantised circular photogalvanic current, the chiral magnetic effect and other novel transport and optics effects forbidden in known topological conductors, such as TaAs \cite{rev6, TaAs1, TaAs2, ARPES-TaAs1, ARPES-TaAs2}.

Incorporating these paradigms into a broader search, we have studied various candidate nonmagnetic and magnetic conductors, such as pyrochlore iridates, MoTe$_2$, WTe$_2$, Mn$_3$Ge, Ge$_3$Sn, Mn$_3$Sn, Na$_3$Bi, GdPtBi, the LuPtSb family, HgCr$_2$Se$_4$, LaPtBi, Co$_3$Sn$_2$S$_2$, Fe$_3$Sn$_2$ and the RhSi family, with advanced spectroscopic techniques. Many of the materials exhibit either large co-existing trivial bulk Fermi surfaces or surface reconstruction masking topological states. And as often is the case with surface-sensitive techniques, the experimentally realised surface potential associated with a cleaved crystal may or may not allow these unusual electronic states to be observed. Thus, despite the new search paradigms, the discovery of topological materials that are suitable for spectroscopic experiments has remained a significant challenge. Of all the materials we explored, we observed that the $X$Si ($X=$ Co, Rh) family of chiral crystals comes close to the experimental realization of the sought-after ideal topological conductor. Additionally it is a novel type of topological conductor beyond Weyl semimetals. Here we report high-resolution angle-resolved photoemission spectroscopy (ARPES) measurements in combination with state-of-the-art \textit{ab initio} calculations to demonstrate novel topological chiral properties in CoSi and RhSi. These chiral crystals approach ideal topological conductors because of their large Fermi arcs which is related to the fact that they host the minimum non-zero number of chiral fermions---topological properties which we experimentally visualise for the first time.

\begin{figure}
\centering
\includegraphics[width=165mm]{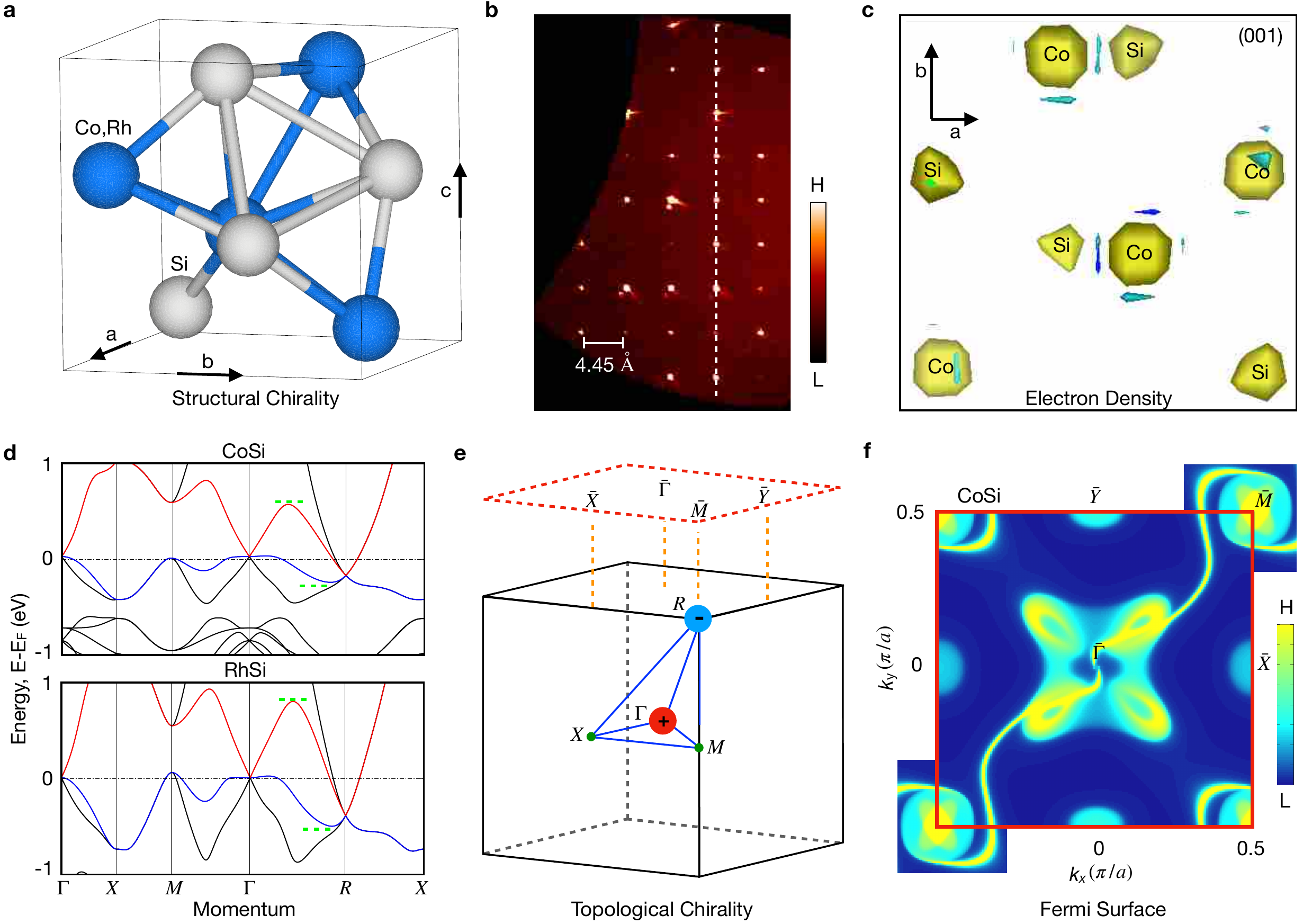}
\caption{\textbf{Structural chirality and topological chirality in CoSi and RhSi.}
\textbf{a}, Chiral crystal structure of $X$Si ($X=$ Co, Rh), space group $P2_{1}3$, No. 198. \textbf{b}, Single crystal X-ray diffraction precession pattern of the (0$kl$) planes of CoSi at 100 K. The resolved spots confirm space group $P2_{1}3$ with lattice constant $a=4.433(4)\ \textrm{\AA}$. \textbf{c}, Three-dimensional Fourier map showing the electron density in the $B20$ CoSi structure. \textbf{d}, \textit{Ab initio} calculation of the electronic bulk band structure along high-symmetry lines. A 3-fold degenerate topological chiral fermion is predicted at $\Gamma$ and a 4-fold topological chiral fermion at $R$; these carry Chern numbers $+2$ and $-2$, respectively. The highest valence (blue) and lowest conduction (red) bands fix a topologically non-trivial energy window (green dotted lines). \textbf{e}, Bulk Brillouin zone (BZ) and (001) surface BZ with high-symmetry points and the predicted chiral fermions (red and blue spheres) marked. \textbf{f}, \textit{Ab initio} calculation of the surface spectral weight on the (001) surface for CoSi, with the (001) surface BZ marked (red box). The predicted bulk chiral fermions project onto $\bar{\Gamma}$ and $\bar{M}$, connected by a pair of topological Fermi arcs extending diagonally across the surface BZ.
}
\label{Fig1}
\end{figure}

The $X$Si ($X=$ Co, Rh) family of materials crystallises in a structurally chiral cubic lattice with space group $P2_{1}3$, No. 198 (Fig.~\ref{Fig1}\pana). We confirmed the chiral crystal structure of our CoSi samples by single crystal X-ray diffraction (XRD; Fig.~\ref{Fig1}\panb), with associated 3D Fourier map (Fig.~\ref{Fig1}\panc). We found a Flack factor of $\sim91\%$, which indicates that our samples are predominantly of a single structural chirality. \textit{Ab initio} electronic bulk band structure calculations predict that both chiral crystals exhibit a 3-fold degeneracy at $\Gamma$ near the Fermi level, \ef\ (Fig.~\ref{Fig1}\pand). This degeneracy is described by a low-energy Hamiltonian which exhibits a 3-fold fermion associated with Chern number $+2$ \cite{RhSi, CoSi}. We refer to this Chern number as a chiral charge, a usage of the term ``chiral'' which is distinct from the notion of structural chirality defined above and which also motivates our use of the term ``chiral fermion''  to describe these topological band crossings. The $R$ point hosts a 4-fold degeneracy corresponding to a 4-fold fermion with Chern number $-2$. These two higher-fold chiral fermions are pinned to opposite TRIMs and are consequently constrained to be maximally separated in momentum space (Fig.~\ref{Fig1}\pane), suggesting that CoSi might provide a near-ideal platform for accessing topological phenomena using a variety of techniques. The hole pocket at $M$ is topologically trivial at the band relevant for low-energy physics, but it is well-separated in momentum space from the $\Gamma$ and $R$ topological crossings. It is not expected to affect topological transport, such as the chiral anomaly (note that $\sigma_{xx}$ is not topological transport). The two higher-fold chiral fermions lead to a net Chern number of zero in the entire bulk Brillouin zone (BZ), as expected from theoretical considerations \cite{NielsenNinomiya1}. CoSi and RhSi also satisfy a key criterion for an ideal topological conductor, namely that they have only two chiral fermions in the bulk BZ, the minimum non-zero number allowed.

The two chiral fermions remain topologically non-trivial over a wide energy range. In particular, CoSi maintains constant-energy surfaces with non-zero Chern number over an energy window of 0.85 eV, while for RhSi this window is 1.3 eV (Fig.~\ref{Fig1}\pand). This prediction suggests that $X$Si satisfies another criterion for an ideal topological conductor---a large topologically non-trivial energy window. An \textit{ab initio} calculated Fermi surface shows that the projection of the higher-fold chiral fermions to the (001) surface results in a hole (electron) pocket at $\bar{\Gamma}$ ($\bar{M}$) with Chern number $+2$ ($-2$; Fig.~\ref{Fig1}\panf). As a result, we expect that $X$Si hosts Fermi arcs of length $\pi$ spanning the entire surface BZ, again suggesting that these materials may realise a near-ideal topological conductor.

\begin{figure}[t]
\centering
\includegraphics[width=165mm]{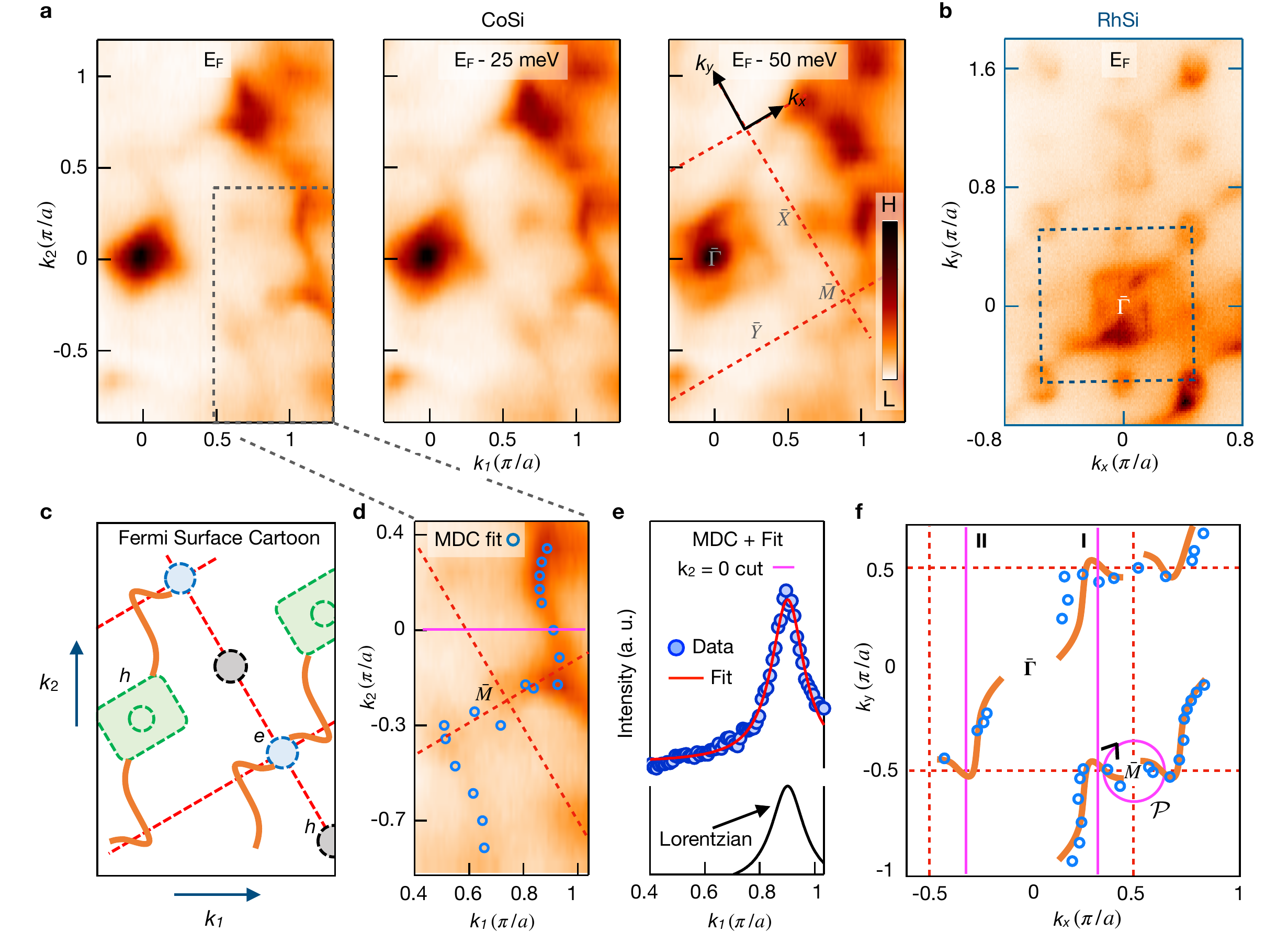}
\caption{\textbf{Fermiology of the (001) surface electronic structure in CoSi and RhSi.}
\textbf{a}, ARPES constant-energy contours for CoSi measured at incident photon energy 50 eV with the Brillouin zone (BZ) boundary marked (rightmost panel, red dotted line). We observe long winding states connecting the $\bar{\Gamma}$ and $\bar{M}$ pockets. \textbf{b}, Fermi surface for RhSi measured at incident photon energy 82 eV with BZ boundary marked (blue dotted line). Again, we observe long winding states extending diagonally across the BZ (Extended Data Fig.~\ref{ExtFig1}). \textbf{c}, Schematic of the measured Fermi surface for CoSi showing hole-like ($h$) and electron-like ($e$) bulk pockets and long winding states (orange). \textbf{d}, Zoomed-in view of the long winding states with a trajectory obtained by fitting Lorentzians to the momentum distribution curves (MDCs) of the ARPES spectrum (blue circles). \textbf{e}, Representative Lorentzian fit to the MDC along $k_1$ for $k_2 = 0$ at binding energy $E-E_\textrm{F}=-10$ meV (Extended Data Fig.~\ref{ExtFig5}). \textbf{f}, Schematic overlaid with Lorentzian peaks extracted from the MDCs passing through the long winding states. We mark two straight cuts, Cut I and II (magenta arrows), as well as a closed loop cut $\mathcal{P}$ (magenta circle).
}
\label{Fig2}
\end{figure}

Using low-photon-energy ARPES, we experimentally study the (001) surface of CoSi and RhSi to reveal their surface electronic structure. For CoSi, the measured constant-energy contours show the following dominant features: two concentric contours around the $\bar{\Gamma}$ point, a faint contour at the $\bar{X}$ point, and long winding states extending along the $\bar{M}-\bar{\Gamma}-\bar{M}$ direction (Fig.~\ref{Fig2}\pana). Both the $\bar{\Gamma}$ and $\bar{X}$ pockets show a hole-like behaviour (Extended Data Fig.~\ref{ExtFig8_bulk_pockets}). The measured surface electronic structure for RhSi shows similar features (Fig.~\ref{Fig2}\panb; Extended Data Fig.~\ref{ExtFig1}). Using only our spectra, we first sketch the key features of the experimental Fermi surface for CoSi (Fig.~\ref{Fig2}\panc). Then, to better understand the $\textit{k}$-space trajectory of the long winding states in CoSi, we study Lorentzian fits to the momentum distribution curves (MDCs) of the ARPES spectrum. We plot the Lorentzian peak positions as the extracted band dispersion (Fig.~\ref{Fig2}\pand, \pane) and we find that the long winding states extend from the center of the BZ to the $\bar{M}$ pocket (Fig.~\ref{Fig2}\panf). To better understand the nature of these states, we perform an ARPES photon energy dependence and we find that the long winding states do not disperse as we vary the photon energy, suggesting that they are surface states (Extended Data Fig.~\ref{ExtFig6}). Moreover, we observe an overall agreement between the ARPES data and the $\textit{ab initio}$ calculated Fermi surface, where topological Fermi arcs connect the $\bar{\Gamma}$ and $\bar{M}$ pockets (Fig.~\ref{Fig1}\panf). Taken together, these results suggest that the long winding states observed in ARPES may be topological Fermi arcs.

\begin{figure}[t]
\includegraphics[width=165mm]{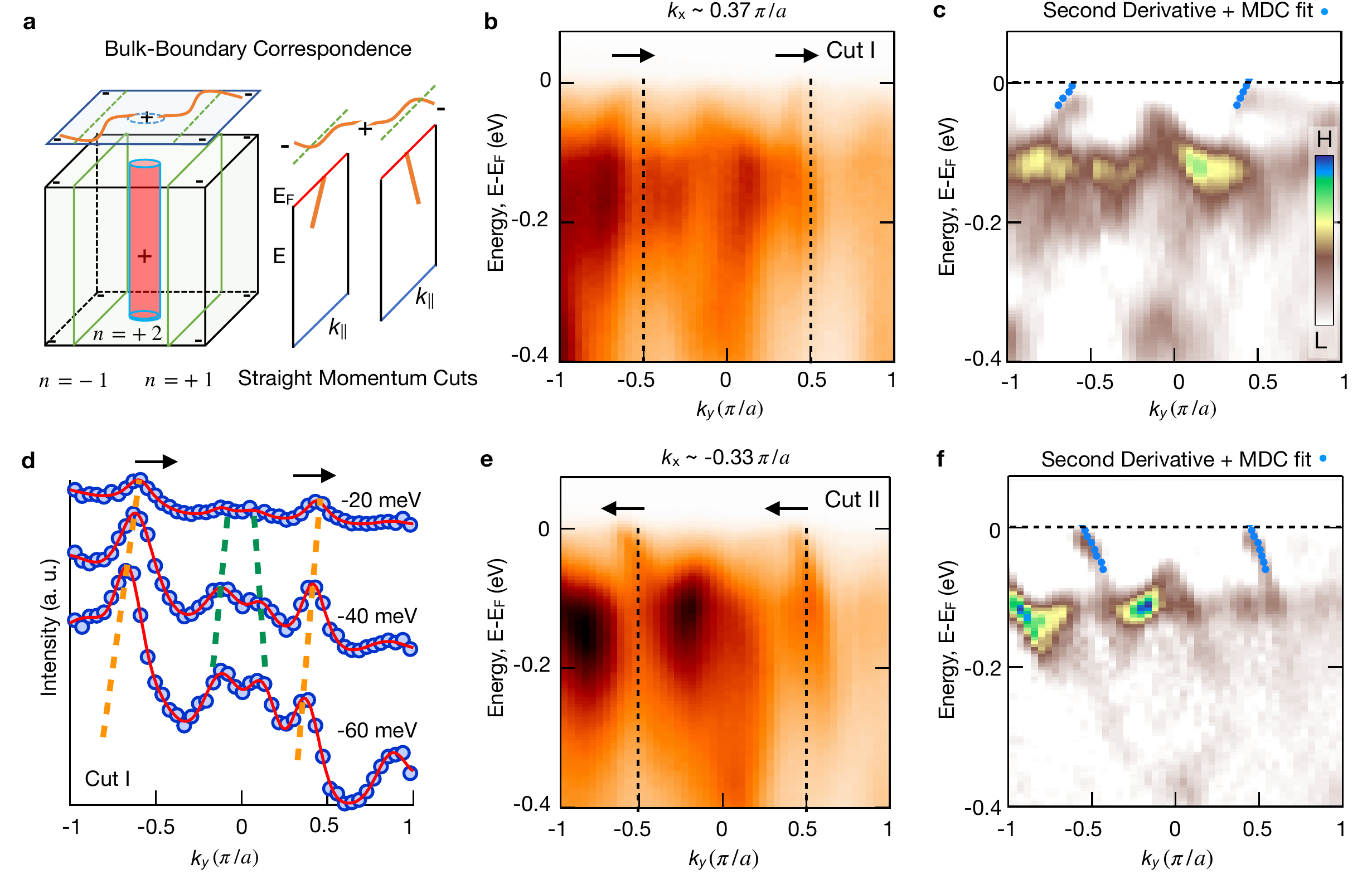}
\caption{ \textbf{Observation of topological chiral edge modes in CoSi.}
\textbf{a}, Left panel: schematic 3D bulk and 2D surface Brillouin zone (BZ) of CoSi with chiral fermions marked ($\pm{}$) and including several examples of 2D manifolds hosting a Chern number $n$ (green planes, red cylinder) \cite{arcDetect1}. Every plane in the bulk has a non-zero $n$. The cylinder enclosing the bulk chiral fermion at $\Gamma$ has $n=+2$. Right panel: Fermi arcs (orange curves) show up as chiral edge modes (orange lines). Energy-momentum cuts on opposite sides of $\bar{\Gamma}$ are expected to show chiral modes propagating in opposite directions. \textbf{b}, ARPES spectrum along Cut I (as marked in Fig.~\ref{Fig2}f), suggesting two right-moving chiral edge modes (black arrows). Vertical dotted lines mark the BZ boundaries. \textbf{c}, Second derivative plot of Cut I, with fitted Lorentzian peaks for a series of momentum distribution curves (MDCs) to track the dispersion (blue dots). \textbf{d}, MDCs and Lorentzian fits tracking the chiral edge modes (dotted orange lines) for Cut I. \textbf{e}, Same as (b), but for Cut II, suggesting two left-moving chiral edge modes (black arrows). \textbf{f}, Same as (c), but for Cut II. The difference in the net number of chiral edge modes on Cut I and Cut II suggests a Chern number $+2$ living near $\bar{\Gamma}$.
}
\label{Fig3}
\end{figure}

Grounded in the framework of topological band theory, the bulk-boundary correspondence of chiral fermions makes it possible for ARPES (spectroscopic) measurements to determine the Chern numbers of a crystal by probing the surface state dispersion (Fig.~\ref{Fig3}\pana; Methods). Such spectroscopic methods to determine Chern numbers have become well-accepted in the field \cite{Nobel}. Using such approach and its spectroscopic analogs, we provide two spectral signatures of Fermi arcs in CoSi. We first look at the dispersion of the candidate Fermi arcs along a pair of energy-momentum cuts on opposite sides of the $\bar{\Gamma}$ pocket, taken at fixed $+k_x$ (Cut I) and $-k_x$ (Cut II; Fig.~\ref{Fig2}\panf). In Cut I, we observe two right-moving chiral edge modes (Fig.~\ref{Fig3}\panb,\panc). Since the cut passes through two BZs (Fig.~\ref{Fig2}\panf), we associate one right-moving mode with each BZ. Next, we fit Lorentzian peaks to the MDCs and we find that the extracted dispersion again suggests two chiral edge modes (Fig.~\ref{Fig3}\pand). Along Cut II, we observe two left-moving chiral edge modes (Fig.~\ref{Fig3}\pane,\panf). Consequently, one chiral edge mode is observed for each measured surface BZ on Cuts I and II, but with opposite Fermi velocity direction. In this way, our ARPES spectra suggest that the number of chiral edge modes $n$ changes by $+2$ when the $k$-slice is swept from Cut I to Cut II. This again suggests that the long winding states are topological Fermi arcs. Moreover, these ARPES results imply that projected topological charge with net Chern number $+2$ lives near $\bar{\Gamma}$.

Next we search for other Chern numbers encoded by the surface state band structure. We study an ARPES energy-momentum cut on a loop $\mathcal{P}$ enclosing $\bar{M}$ (Fig.~\ref{Fig4}a, inset). Again following the bulk-boundary correspondence, we aim to extract the Chern number of chiral fermions projecting on $\bar{M}$. The cut $\mathcal{P}$ shows two right-moving chiral edge modes dispersing towards \ef\ (Fig.~\ref{Fig4}\pana,\panb), suggesting a Chern number $-2$ on the associated bulk manifold. Furthermore, the \textit{ab initio} calculated surface spectral weight along $\mathcal{P}$ is consistent with our experimental results (Fig.~\ref{Fig4}\panc). Our ARPES spectra on Cut I, Cut II and $\mathcal{P}$ suggest that CoSi hosts a projected chiral charge of $+2$ at $\bar{\Gamma}$ with its partner chiral charge of $-2$ projecting on $\bar{M}$. This again provides evidence that the long winding states are a pair of topological Fermi arcs which traverse the surface BZ on a diagonal, connecting the $\bar{\Gamma}$ and $\bar{M}$ pockets. Our ARPES spectra on RhSi also provide evidence for gigantic topological Fermi arcs following a similar analysis (Extended Data Fig.~\ref{ExtFig1}).

To further explore the topological properties of CoSi, we examine in greater detail the structure of the Fermi arcs near $\bar{M}$. We consider the dispersion on $\mathcal{P}$ (plotted as a magenta loop in Fig.~\ref{Fig4}d, inset) and we also extract a dispersion from Lorentzian fitting on a second, tighter circle (black loop; Extended Data Fig.~\ref{ExtFig_hel_fit}). We observe that as we decrease the binding energy (approach \ef), the extracted dispersion spirals in a clockwise fashion on both loops, suggesting that as a given $\textit{k}$ point traverses the loop, the energy of the state does not return to its initial value after a full cycle. Such a non-trivial electronic dispersion directly signals a projected chiral charge at $\bar{M}$ (Fig.~\ref{Fig4}d). In fact, the extracted dispersion is characteristic of the helicoid structure of topological Fermi arcs as they wind around a chiral fermion (Fig.~\ref{Fig4}e), suggesting that CoSi provides a rare example of a non-compact surface in nature \cite{HelicodalFermiArcs, KramersWeyl}.

To further understand these experimental results, we consider the \textit{ab initio} calculated spectral weight for the (001) surface and we observe a pair of Fermi arcs winding around the $\bar{\Gamma}$ and $\bar{M}$ pockets in a counterclockwise and clockwise manner, respectively, with decreasing binding energy (approaching \ef; Fig.~\ref{Fig4}\panf). The clockwise winding around $\bar{M}$ is consistent with our observation by ARPES of a $-2$ projected chiral charge. Moreover, from our \textit{ab initio} calculations, we predict that the $-2$ charge projecting to $\bar{M}$ arises from a 4-fold chiral fermion at the bulk $R$ point (Fig.~\ref{Fig1}d). The $+2$ chiral charge which we associate with $\bar{\Gamma}$ from ARPES (Fig.~\ref{Fig3}) is further consistent with the 3-fold chiral fermion predicted at the bulk $\Gamma$ point. By fully accounting for the predicted topological charges in experiment, our ARPES results suggest the demonstration of a topological chiral crystal in CoSi. We can similarly account for the predicted topological charges in RhSi from our ARPES data (Extended Data Fig.~\ref{ExtFig1},~\ref{ExtFig2}).

The surface state dispersions in our ARPES spectra, taken together with the topological bulk-boundary correspondence established in theory \cite{Wan, arcDetect1}, demonstrate that CoSi is a topological chiral crystal. This experimental result is further consistent with the numerical result determined from first-principles calculations of the surface state dispersions and topological invariants. Unlike previously-reported WSMs, the Fermi arcs which we observe in CoSi and RhSi stretch diagonally across the entire (001) surface Brillouin zone, from $\bar{\Gamma}$ to $\bar{M}$. In fact, the Fermi arcs in $X$Si are longer than those found in TaAs by a factor of thirty. Our surface band structure measurements also demonstrate two well-separated Fermi pockets carrying Chern number $\pm{2}$. Lastly, we observe for the first time in an electronic material the helicoid structure of topological Fermi arcs, offering an example of a non-compact  surface \cite{HelicodalFermiArcs, KramersWeyl}. Our results suggest that CoSi and RhSi are excellent candidates for studying topological phenomena distinct to chiral fermions, using a variety of techniques \cite{rev4, rev7, rev6}.


Crucial for applications, the topologically non-trivial energy window in CoSi is an order of magnitude larger than that in TaAs \cite{ARPES-TaAs1, ARPES-TaAs2}, rendering its quantum properties robust against changes in surface chemical potential and disorder. Moreover, the energy offset between the higher-fold chiral fermions at $\Gamma$ and $R$ is predicted to be $\sim225$ meV. Such an energy offset is essential for inducing the chiral magnetic effect and its optical analog \cite{chiralmagenticeffect1} and the quantised photogalvanic effect (optical) \cite{chiral_photogalvanic}. When coupled to a compatible superconductor, CoSi is a compelling platform for studying the superconducting pairing of Fermi surfaces with non-zero Chern numbers, which may be promising for realizing a new type of topological superconducting phase recently proposed by Li and Haldane \cite{Haldane} which can be probed with STM-based spectroscopy. CoSi further opens the door to exploring other exotic quantum phenomena when combined with the isochemical material FeSi. Fe$_{1-x}$Co$_x$Si may simultaneously host $\textit{k}$-space topological defects (chiral fermions) and real-space topological defects (skyrmions) and their interplay which can also be probed by STM/STS. Through our observation of a helicoid surface state and its topological properties, our results suggest the discovery of the first structurally chiral crystal that is also topological. In this way, our work provides a much-needed new and next-generation platform for further study and search for novel types of topological conductors.

\ \\


\clearpage

\clearpage

\clearpage

\begin{figure}
\includegraphics[width=155mm]{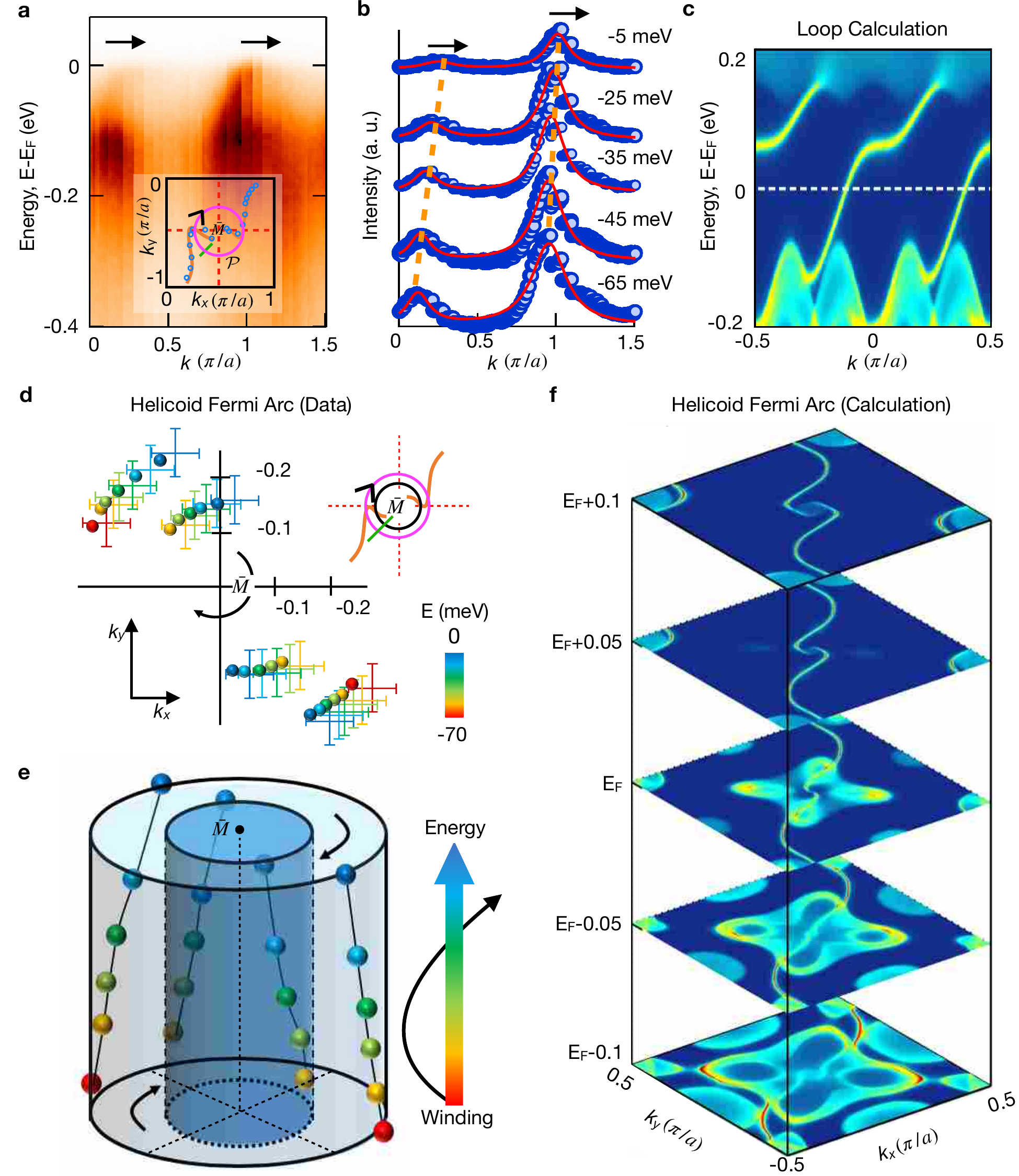}
\end{figure}
\begin{figure*}[t!]
\caption{\textbf{Observation of helicoid Fermi arcs.}
\textbf{a}, ARPES spectrum along loop $\mathcal{P}$, revealing two right-moving chiral edge modes and a projected chiral charge of $-2$ at $\bar{M}$. Inset: definition of the loop $\mathcal{P}$, starting from the green mark and proceeding clockwise. \textbf{b}, Lorentzian fits to a series of momentum distribution curves (MDCs) along $\mathcal{P}$ to track the dispersion of the chiral edge modes. \textbf{c}, \textit{Ab initio} calculation of the dispersion along a loop around $\bar{M}$ showing two right-moving chiral edge modes, consistent with the ARPES data. \textbf{d}, Extracted dispersion of the chiral edge modes on $\mathcal{P}$ and a second inner loop from Lorentzian fits to the MDCs. Error bars correspond to the momentum resolution. Inset: definition of the second inner loop (black). We observe that the chiral edge modes spiral in a clockwise way with decreasing binding energy (approaching \ef). \textbf{e}, Perspective plot of (d), where the two loops now correspond to two concentric cylinders. The winding of the chiral modes around $\bar{M}$ as a function of binding energy suggests that the Fermi arcs have a helicoid structure \cite{HelicodalFermiArcs, KramersWeyl}. \textbf{f}, \textit{Ab initio} calculated constant-energy contours, consistent with the helicoid Fermi arc structure observed in our ARPES spectra.}
\label{Fig4}
\end{figure*}
 \clearpage

\textbf{\large Supplementary Information: Topological chiral crystals with helicoid arc states}

\textbf{\textit{Observation of Fermi arcs in RhSi.}} Motivated by an interest in establishing families of closely-related topological materials \cite{burkov, Jia, H.Weyl, typeIIWeyl, TypeIIWeyl1, TypeIIWeyl2, TypeIIWeyl3, TypeIIWeyl4}, we expand our experiment to include RhSi, an isoelectronic cousin of CoSi. Rhodium silicide, RhSi, crystallises in a chiral cubic lattice, space group $P2_{1}3$, No. 198. The calculated electronic band structure is generally similar to that of CoSi (side-by-side comparison in Fig.~\ref{ExtFig1}\pand). Constant-energy contours measured by ARPES further suggest features similar to those in CoSi (Extended Data Fig.~\ref{ExtFig1}\pana). In particular, we observe contours at the $\bar{\Gamma}$, $\bar{X}$ and $\bar{Y}$ points and long winding states that extend diagonally from $\bar{\Gamma}$ to $\bar{M}$. Again taking advantage of the bulk-boundary correspondence, we focus on counting chiral edge modes to determine the topological nature of RhSi (Extended Data Fig.~\ref{ExtFig1}\panb,\panc). A second-derivative plot of the Fermi surface further suggests the presence of long states stretching diagonally across the BZ, motivating a study of energy-momentum cuts through these states (Extended Data Fig.~\ref{ExtFig1}\pand-\pang). The cuts show that the long states take the form of a right-moving chiral edge mode (Cut I) and a left-moving chiral edge mode (Cut II) on opposite sides of $\bar{\Gamma}$. The net difference in the number of right-moving chiral edge modes suggests that a Chern number of $+2$ projects to $\bar{\Gamma}$ and that the long states are topological Fermi arcs. Proceeding again by analogy to CoSi, we study the band structure along a loop $\mathcal{M}$ enclosing the $\bar{M}$ point and we observe two right-moving chiral edge modes, suggesting a Chern number of $-2$ at the $R$ point. These results suggest that RhSi, an isoelectronic cousin of CoSi, provides another example of a near-ideal topological conductor.

\textbf{\textit{Helicoid structure of the Fermi arcs in RhSi.}} We study the Fermi arcs near $\bar{M}$ in RhSi, by analogy with the analysis performed for CoSi (Fig.~\ref{Fig4}). First, we consider ARPES energy-momentum cuts on an inner and outer loop enclosing the $\bar{M}$ point and we observe signatures of two right-moving chiral edge modes on each loop, suggesting an enclosed projected chiral charge of $-2$ (Extended Data Fig.~\ref{ExtFig2}\pana-\pand). We further fit Lorentzians to the MDCs and we find that the extracted dispersion exhibits a clockwise winding on both loops for decreasing binding energy (approaching \ef; Extended Data Fig.~\ref{ExtFig2}\pane,\panf). This dispersion again suggests a chiral charge $-2$ at $\bar{M}$ and is characteristic of the helicoid structure of Fermi arcs as they wind around a chiral fermion. Lastly, we study $\textit{ab initio}$ calculations of the constant-energy contours (Extended Data Fig.~\ref{ExtFig2}\pang) which exhibit the same winding pattern. These ARPES results suggest that RhSi, like CoSi, exhibits helicoid topological Fermi arcs.

\textbf{\textit{Bulk-boundary correspondence in CoSi and RhSi.}} By measuring the surface state band structure of a crystal, ARPES is capable of demonstrating Fermi arcs and counting Chern numbers \cite{arcDetect1}. We briefly review the details of this method. First, recall that topological invariants are typically defined for a gapped system. A topological conductor is by definition gapless and so we cannot define a topological invariant for the full bulk Brillouin zone of the three-dimensional system. However, we may be able to choose certain two-dimensional $\textit{k}$-space manifolds where the band structure is fully gapped and a two-dimensional invariant can then be defined on this slice of momentum space (Fig.~\ref{Fig3}a). We consider specifically the case of a chiral fermion in a topological conductor and we study a gapped momentum-space slice which cuts in between two chiral fermions. By definition, a chiral fermion is associated with a non-zero Chern number, so at least some of these slices will be characterized by a non-zero Chern number. Following the bulk-boundary correspondence, these slices will then contribute chiral edge modes to the surface state dispersion. If we image collecting together the chiral edge modes from all of the gapped slices, we will assemble the entire topological Fermi arc of the topological conductor. If we run the bulk-boundary correspondence in reverse, we can instead measure the surface states by ARPES and count the chiral edge modes on a particular one-dimensional slice of the surface Brillouin zone to determine the Chern number of the underlying two-dimensional slice of the bulk Brillouin zone. These Chern numbers in turn fix the chiral charges of the topological fermions.

Next we highlight two spectral signatures that can determine the chiral charges specifically in $X$Si ($X=$ Co, Rh). First, we consider chiral edge modes along straight $\textit{k}$-slices. Theoretically, the two-dimensional $\textit{k}$-slices on the two sides of the chiral fermion at $\Gamma$ are related by time-reversal symmetry and therefore should have equal and opposite Chern numbers ($n_{l}=-n_{r}$; Fig. 3b). On the other hand, because the 3-fold chiral fermion at $\Gamma$ is predicted to carry chiral charge $+2$, we expect that the difference should be $n_{l}-n_{r}=+2$, resulting in $n_{l(r)}=\pm{1}$. Therefore we expect one net left-moving chiral edge mode on one cut and one net right-moving chiral edge mode on the other. For the second signature, we study the chiral edge modes along a closed loop in the surface BZ. Any loop that encloses the projected chiral charge of $+2$ at $\bar{\Gamma}$ or $-2$ at $\bar{M}$ has this number of net chiral edge modes along this path. By taking advantage of this correspondence, we can count surface states on loops in our ARPES spectra to determine enclosed projected chiral charges.

\textbf{\textit{Growth of CoSi and RhSi single crystals.}} Single crystals of CoSi were grown using a chemical vapor transport (CVT) technique. First, polycrystalline CoSi was prepared by arc-melting stoichiometric amounts of Co slices and Si pieces. After being crushed and ground into a powder, the sample was sealed in an evacuated silica tube with an iodine concentration of approximately $0.25$ mg/cm$^3$. The transport reaction took place at a temperature gradient from $1000\degree$C (source) to $1100\degree$C (sink) for two weeks. The resulting CoSi single crystals had a metallic luster and varied in size from $1$ mm to $2$ mm. Single crystals of RhSi were grown from a melt using the vertical Bridgman crystal growth technique at a non-stoichiometric composition. In particular, we induced a slight excess of Si to ensure a flux growth inside the Bridgman ampoule. First, a polycrystalline ingot was prepared by pre-melting the highly pure constituents under an argon atmosphere using an arc furnace. The crushed powder was poured into a custom-designed sharp-edged alumina tube and then sealed inside a tantalum tube again under an argon atmosphere. The sample was heated to $1550\degree$C and then slowly pooled to the cold zone at a rate of $0.8$ mm$/$h. Single crystals on average $\sim15$ mm in length and $\sim6$ mm in diameter were obtained.

\textbf{\textit{X-ray diffraction.}} Single crystals of CoSi were mounted on the tips of Kapton loops. The low-temperature (100 K) intensity data was collected on a Bruker Apex II X-ray diffractometer with Mo radiation K$\alpha_{1}$ ($\lambda=0.71073\textrm{\AA}^{-1}$). Measurements were performed over a full sphere of $\textit{k}$-space with 0.5$\degree$ scans in $\omega$ with an exposure time of 10 seconds per frame (Extended Data Fig.~\ref{ExtFig3}). The SMART software was used for data acquisition. The extracted intensities were corrected for Lorentz and polarization effects with the SAINT program. Numerical absorption corrections were accomplished with XPREP, which is based on face-indexed absorption \cite{XRay1}. The twin unit cell was tested. With the SHELXTL package, the crystal structures were solved using direct methods and refined by full-matrix least-squares on F$^{2}$ \cite{XRay2}. No vacancies were observed according to the refinement and no residual electron density was detected, indicating that the CoSi crystals were of high quality.

\textbf{\textit{Sample surface preparation for ARPES.}} For CoSi single crystals, the surface preparation procedure followed the conventional \textit{in situ} mechanical cleaving approach. This cleaving method resulted in a low success rate and the resulting surface typically appeared rough under a microscope. We speculate that the difficulty in cleaving CoSi single crystals may be a result of its cubic structure, strong covalent bonding and lack of a preferred cleaving plane. For RhSi single crystals, their large size allowed them to be mechanically cut and polished along the (001) surface. An \textit{in situ} sputtering and annealing procedure, combined with LEED/RHEED characterization, was used to obtain a clean surface suitable for ARPES measurements. The typical spectral line-width was narrowed and the background signal was reduced for these samples as compared with the mechanically-cleaved CoSi samples.

\textbf{\textit{Angle-resolved photoemission spectroscopy.}} ARPES measurements were carried out at beamlines (BL) 10.0.1 and 4.0.3 at the Advanced Light Source in Berkeley, CA, USA. A Scienta R4000 electron analyser was used at BL 10.0.1 and a Scienta R8000 was used at BL 4.0.3. At both beamlines the angular resolutions was $< 0.2 \degree$ and the energy resolution was better than 20 meV. Samples were cleaved or sputtered/annealed \textit{in situ} and measured under vacuum better than $5 \times 10^{-11}$ Torr at $T < 20$ K.

\textbf{\textit{First-principles calculations.}} Numerical calculations of $X$Si ($X=$ Co, Rh) were performed within the density functional theory (DFT) framework using the OPENMX package and the full potential augmented plane-wave method as implemented in the package WIEN2k \cite{DFT1, DFT2, DFT3}. The generalised gradient approximation (GGA) was used \cite{DFT4}. Experimentally measured lattice constants were used in DFT calculations of material band structures \cite{xraydata}. A $\Gamma$-centered $\textit{k}$-point $10\times10\times10$ mesh was used and spin-orbit coupling (SOC) was included in self-consistent cycles. To generate the (001) surface states of CoSi and RhSi, Wannier functions were generated using the $p$ orbitals of Si and the $d$ orbitals of Co and Rh. The surface states were calculated for a semi-infinite slab by the iterative Green's function method.

\textbf{\textit{Electronic bulk band structure.}} The electronic bulk band structure of CoSi is shown with and without spin-orbit coupling (SOC; Extended Data Fig.~\ref{ExtFig4}). The small energy splitting that arises from introducing SOC is approximately 40 meV, which is negligible compared to the $\sim$1.2 eV topologically non-trivial energy window. Our results suggest that we do not need to consider SOC, either theoretically or experimentally, to demonstrate a chiral charge in CoSi.

\textbf{\textit{Tracking the Fermi arcs by fitting ARPES momentum distribution curves.}} The following procedure was performed to track the Fermi arcs in the surface BZ of CoSi (Extended Data Fig.~\ref{ExtFig5}\pana). At $E=E_\textrm{F}$, MDCs were collected for various fixed $k_y$ values (extending along the region of interest). Each MDC was fitted with a Lorentzian function to pinpoint the $k_x$ value that corresponds to the peak maximum. The MDC fitted peak maximum was then plotted on top of the measured Fermi surface and marked with blue circles. Where necessary, the peak maximum corresponding to the $\bar{X}$ pockets is annotated on each MDC fitting panel (Extended Data Fig.~\ref{ExtFig5}\panb). For the chiral edge modes discussed in the main text and the Extended Data, a similar MDC fitting procedure was used at different binding energies to track the dispersion.

\textbf{\textit{Study of the Fermi arcs with varying photon energy.}} We study an MDC cutting through the Fermi arc as a function of incident photon energy (Extended Data Fig.~\ref{ExtFig6}). We find that within the experimental resolution the Fermi arc shows negligible variation from 80 eV to 110 eV, providing additional evidence that it is indeed a surface state.

\clearpage
\newpage

\renewcommand{\tablename}{\textbf{Extended Data Table}}
\renewcommand\thetable{\arabic{table}}

\begin{table}
\begin{center}
\centering
\textsf{\footnotesize
\begin{tabular}{p{6cm}p{3.5cm}p{3.5cm}}
\hline
CoSi & T = 100(2) K & T = 300(2) K\\
\hline
Scan & 1 & 2 \\
F.W. (g/mol) & 87.02 & 87.02 \\
Space Group; $Z$ & $P2_{1}3$ (No.198); 4 & $P2_{1}3$ (No.198); 4\\
$a (\textrm{\AA})$ & 4.433(4) & 4.4245(16)\\
$V (\textrm{\AA}^{3}$) & 87.1(2) & 86.61(9)\\
Absorption Correction & Numerical & Numerical\\
Extinction Coefficient  & 0.39(9) & 0.5(2)\\
$\theta$ range (deg) & 19.301- 31.714 & 18.740- 31.781\\
No. Reflections; R$_{int}$ & 165; 0.0222 & 167;0.0232\\
No. Independent Reflections & 79 & 77\\
No. Parameters & 9 & 9\\
$R_1$; $wR_2$ (all \textit{I}) & 0.0223; 0.0561 & 0.0533; 0.1209\\
Goodness of fit & 1.123 & 1.142\\
Diffraction peak and hole (e$^{-}/\textrm{\AA}^3$) & 0.637; -0.736 & 1.155; -1.895\\
\hline
\end{tabular}
}
\end{center}
\caption{Single crystal crystallographic data for CoSi at 100(2) and 300(2)K.}
\label{ExtTab1}
\end{table}


\clearpage
\newpage

\renewcommand{\figurename}{EXTENDED DATA FIG.}
\renewcommand\thefigure{\arabic{figure}}
 \setcounter{figure}{0}
\begin{figure}
\centering
\includegraphics[width=165mm]{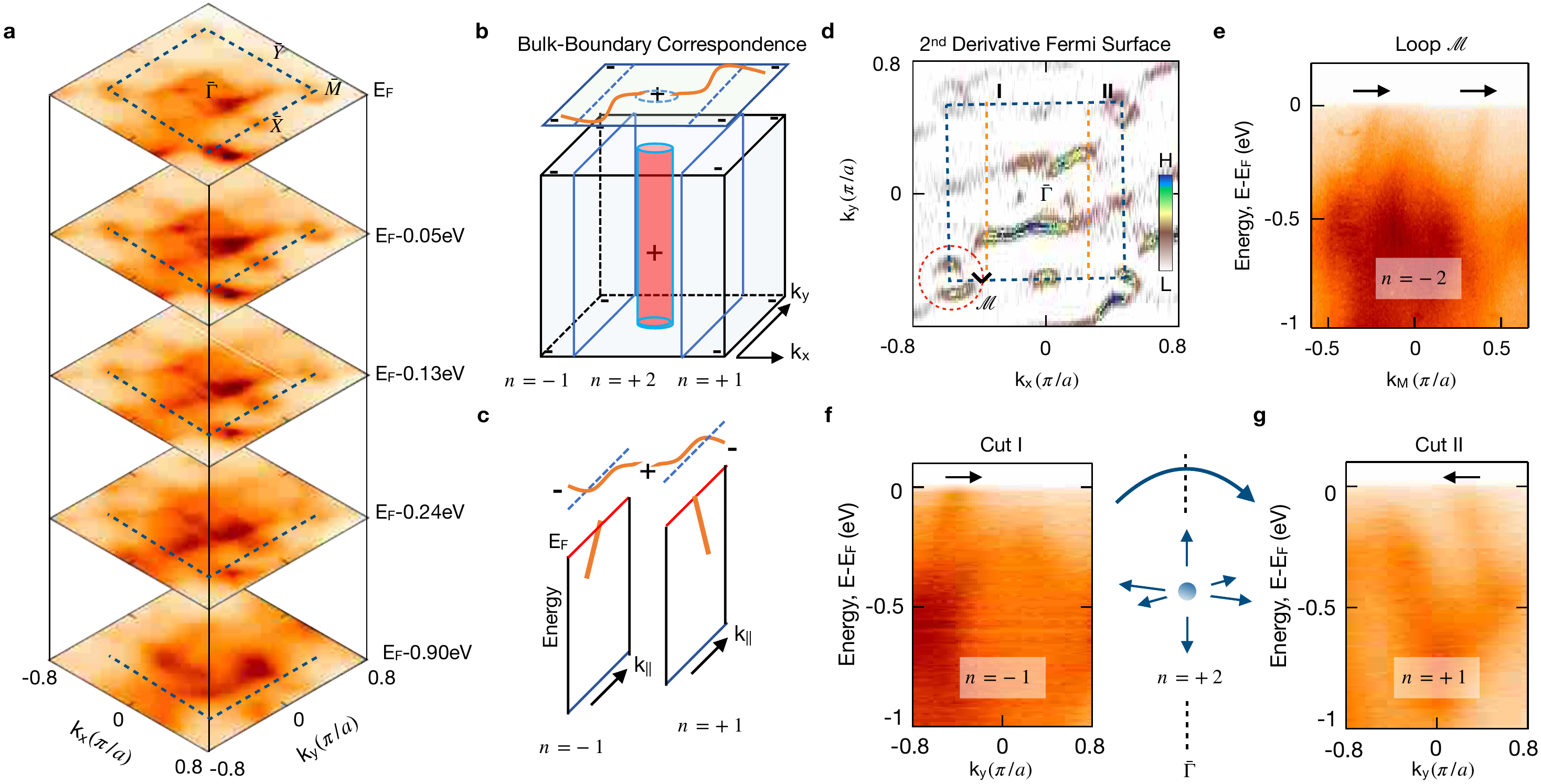}
\caption{\textbf{Topological Fermi arcs in structurally chiral crystal RhSi.}
\textbf{a}, ARPES measured Fermi surface and constant binding energy contours with an incident photon energy of 82eV at 10K. The Brillouin zone (BZ) boundary is marked in blue. \textbf{b}, 3D bulk and 2D surface BZ with higher-fold chiral fermions ($\pm{}$). The planes outlined in blue and the red cylinder are 2D manifolds with indicated Chern number $n$ \cite{arcDetect1}. The cylinder enclosing the bulk chiral fermion at $\Gamma$ has $n=+2$. \textbf{c}, Fermi arcs (orange) connect the projected chiral fermions. Energy dispersion cuts show that these two chiral edge modes are time-reversed partners propagating in opposite directions. \textbf{d}, Second derivative Fermi surface with the straight and loop cuts of interest marked. \textbf{e}, ARPES spectrum along a loop $\mathcal{M}$ showing two right-moving chiral edge modes, suggesting that the $4$-fold chiral fermion at $R$ carries Chern number $-2$. \textbf{f}, ARPES spectrum along Cut I showing a right-moving chiral edge mode. \textbf{g}, ARPES spectrum along Cut II on the opposite side of the $3$-fold chiral fermion at $\bar{\Gamma}$ (illustrated by the blue sphere) showing a left-moving chiral edge mode.
}
\label{ExtFig1}
\end{figure}
\clearpage

\begin{figure}[t]
\centering
\includegraphics[width=165mm]{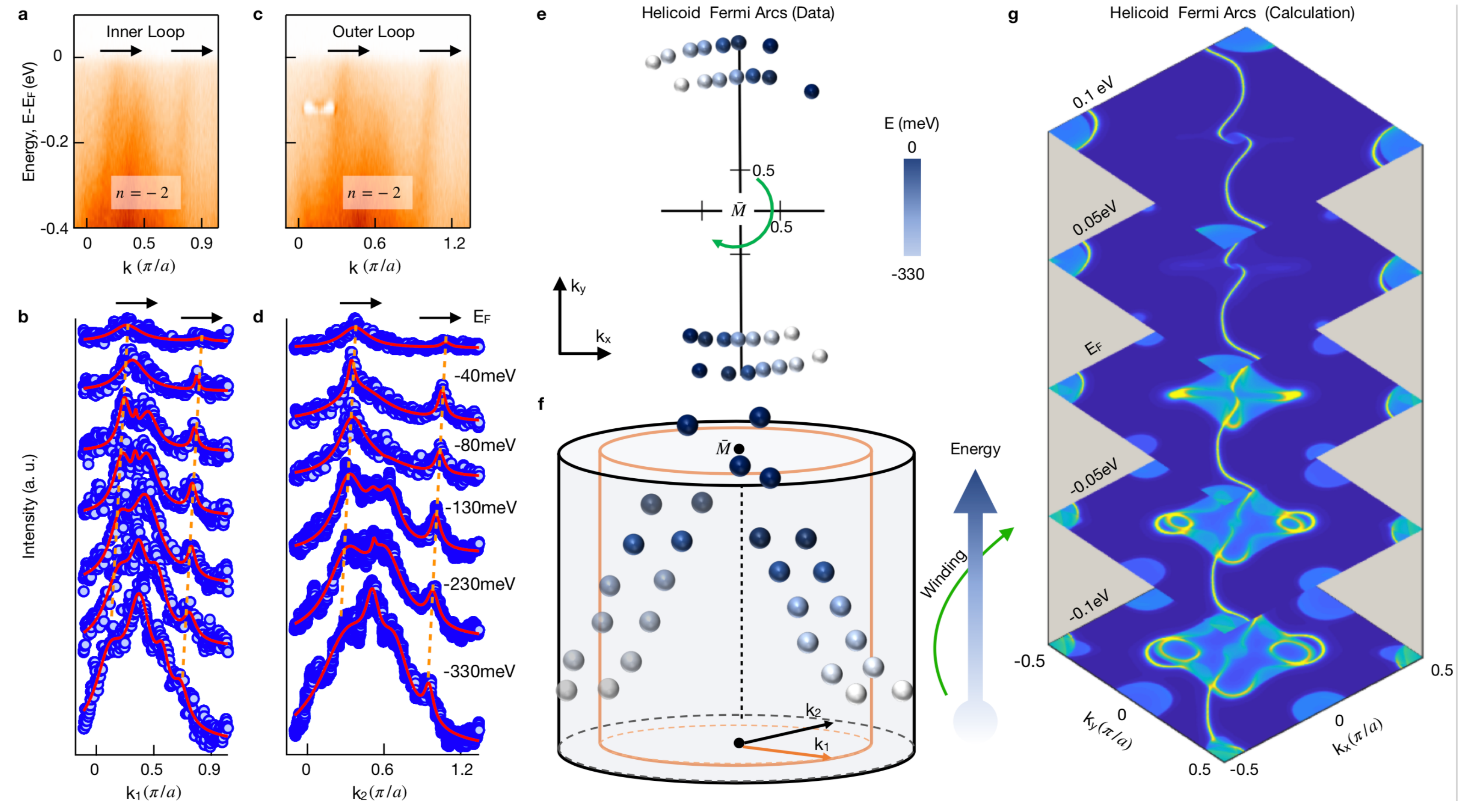}
\caption{\textbf{Fermi arc helicoid in RhSi.} \textbf{a}, Energy dispersion cut on an inner loop of radius $0.18\pi/a$ enclosing the $\bar{M}$. \textbf{b}, Lorentzian fits (red traces) to the momentum distribution curves (MDCs; blue dots) to track the observed chiral edge modes. \textbf{c}, \textbf{d}, Similar analysis to (a, b), but for an outer loop of radius $0.23\pi/a$. Black arrows show the Fermi velocity direction for the chiral edge modes. \textbf{e}, Top view and \textbf{f}, perspective view of the helicoid dispersion extracted from the MDCs, plotted on the two concentric loops, suggesting a clockwise spiral with decreasing binding energy. \textbf{g}, $\textit{Ab initio}$ calculated constant-energy contours show a consistent helicoid structure.
}
\label{ExtFig2}
\end{figure}
\clearpage

\begin{figure}[t]
\includegraphics[width=165mm]{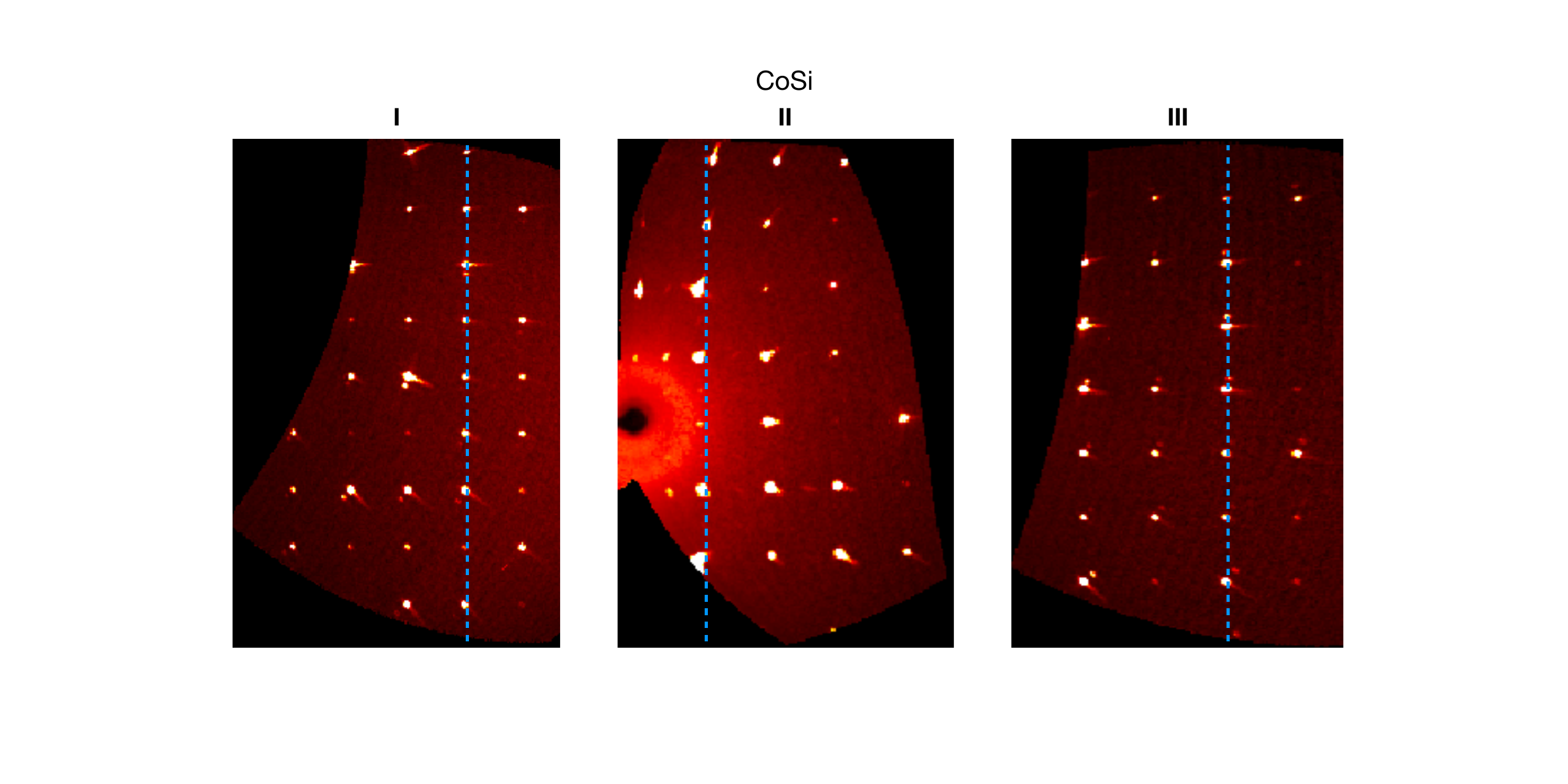}
\caption{ \textbf{X-ray diffraction.} Single crystal X-ray diffraction precession image of the ($0kl$) planes in the reciprocal lattice of CoSi. The resolved spots from scans I and III are consistent with space group $P2_13$ at 100K. This is also shown in the reflection intersection, scan II.
}
\label{ExtFig3}
\end{figure}
\clearpage

\begin{figure}
\includegraphics[width=165mm]{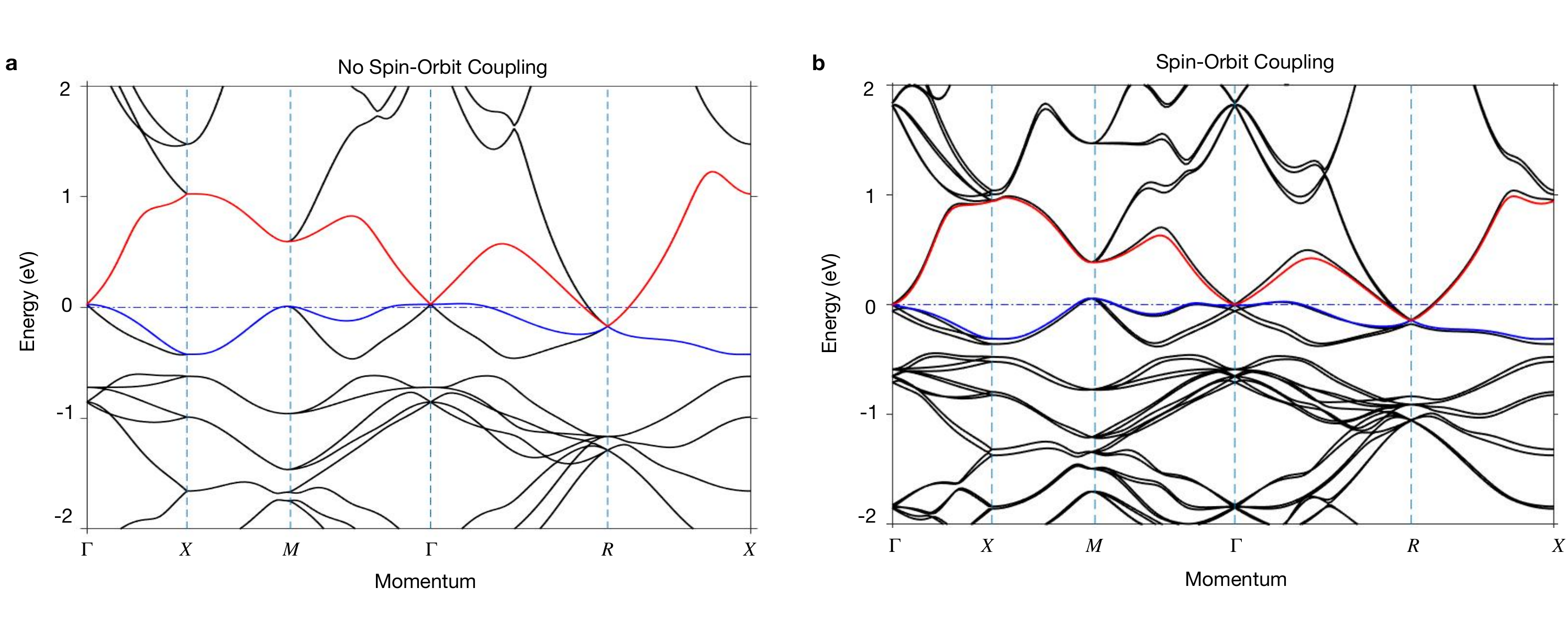}
\caption{\textbf{Electronic bulk band structure.}
\textbf{a}, Band structure of CoSi in the absence of SOC interactions. \textbf{b}, Band structure in the presence of SOC. The highest valence and lowest conduction bands are colored in blue and red, respectively.
}

\label{ExtFig4}
\end{figure}
\clearpage
\newpage

\begin{figure}
\includegraphics[width=165mm]{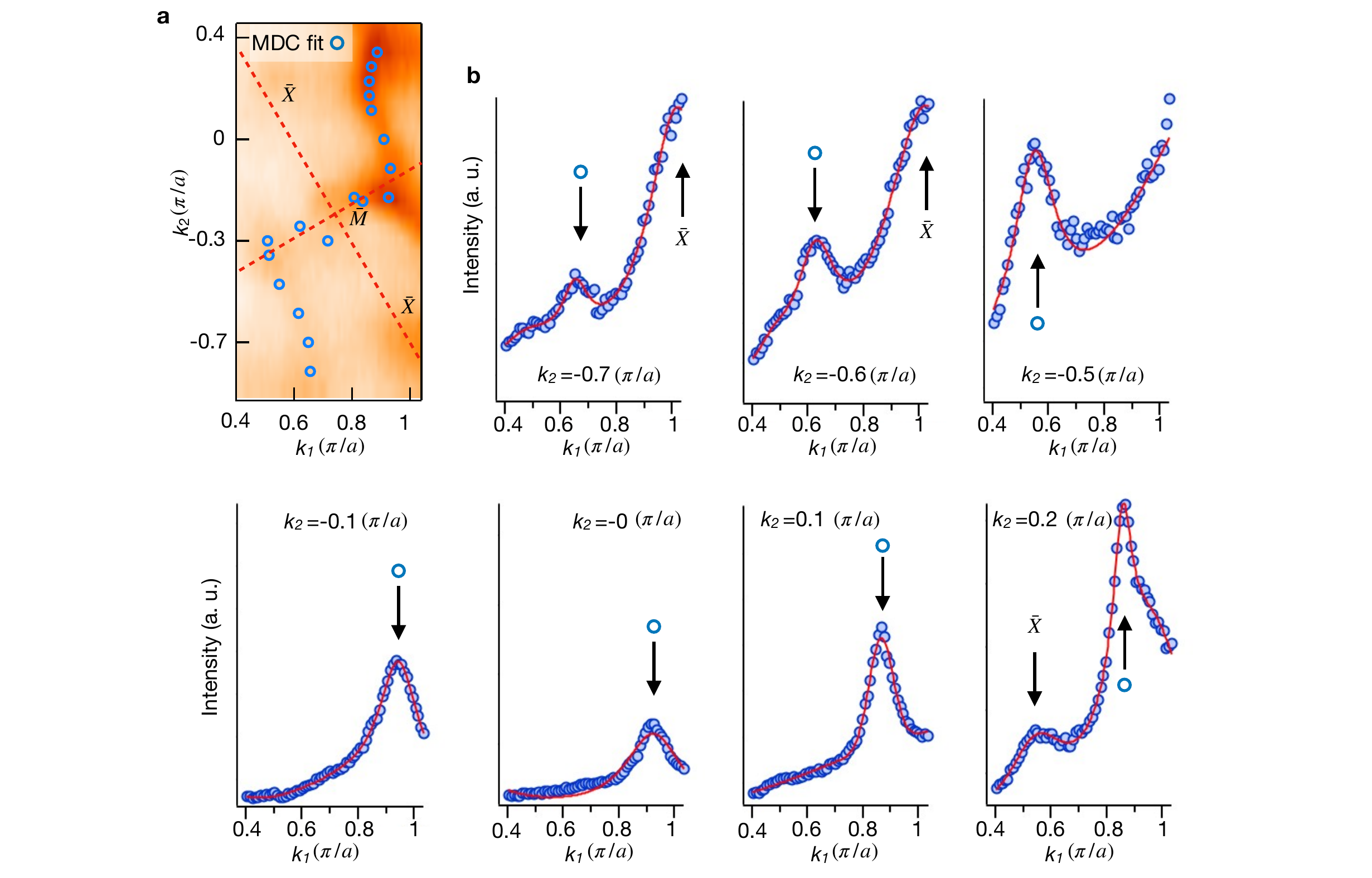}
\caption{\textbf{Tracking the Fermi arcs by fitting Lorentzians to momentum distribution curves (MDCs).}
\textbf{a}, Zoomed-in region of the Fermi surface (Fig.~\ref{Fig2}d), with Fermi arcs tracked (blue circles) and the surface Brillouin zone marked (red dotted lines). \textbf{b}, Representative fits of Lorentzian functions (red lines) to the MDCs (filled blue circles). The peaks indicate the extracted positions of the Fermi arc (open blue circles) and the $\bar{X}$ pocket.
}

\label{ExtFig5}
\end{figure}
\clearpage
\newpage

\begin{figure}
\includegraphics[width=140mm]{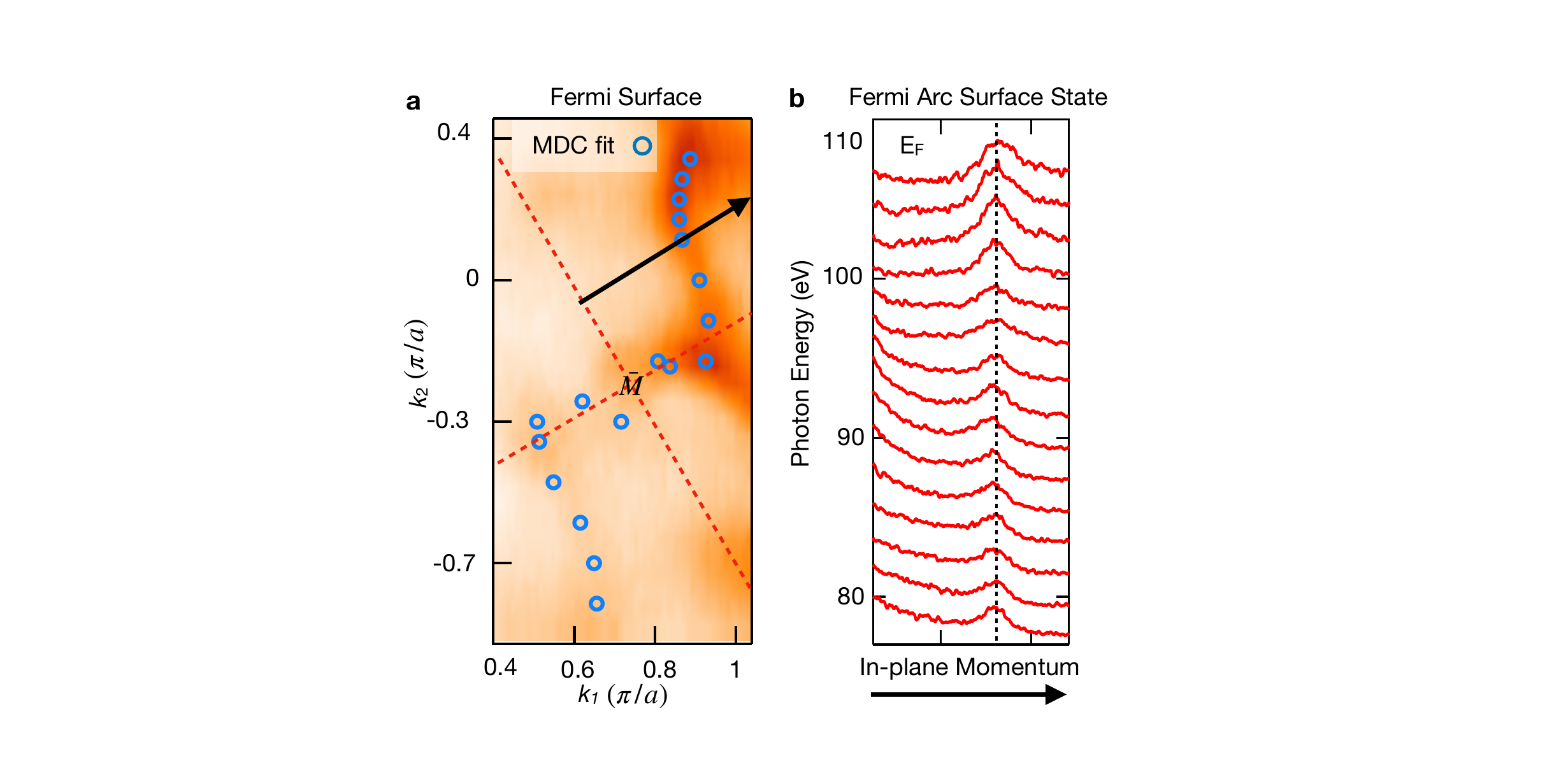}
\caption{\textbf{Photon energy dependence of the Fermi arc in CoSi.} \textbf{a}, Location of the in-plane momentum direction (black arrow) along which a photon energy dependence study was performed, plotted on the Fermi surface (Fig.~\ref{Fig2}d), with surface Brillouin zone (red dotted lines). \textbf{b}, Momentum distribution curves (MDCs) at \ef\ along the in-plane momentum direction illustrated in (a), obtained for a series of photon energies from 80 eV to 110 eV in steps of 2 eV. The peak associated with the Fermi arc shows negligible variation in photon energy (dotted line) within experimental resolution, providing further evidence that the observed Fermi arc is indeed a surface state.
}

\label{ExtFig6}
\end{figure}
\clearpage
\newpage

\begin{figure}
\includegraphics[width=165mm]{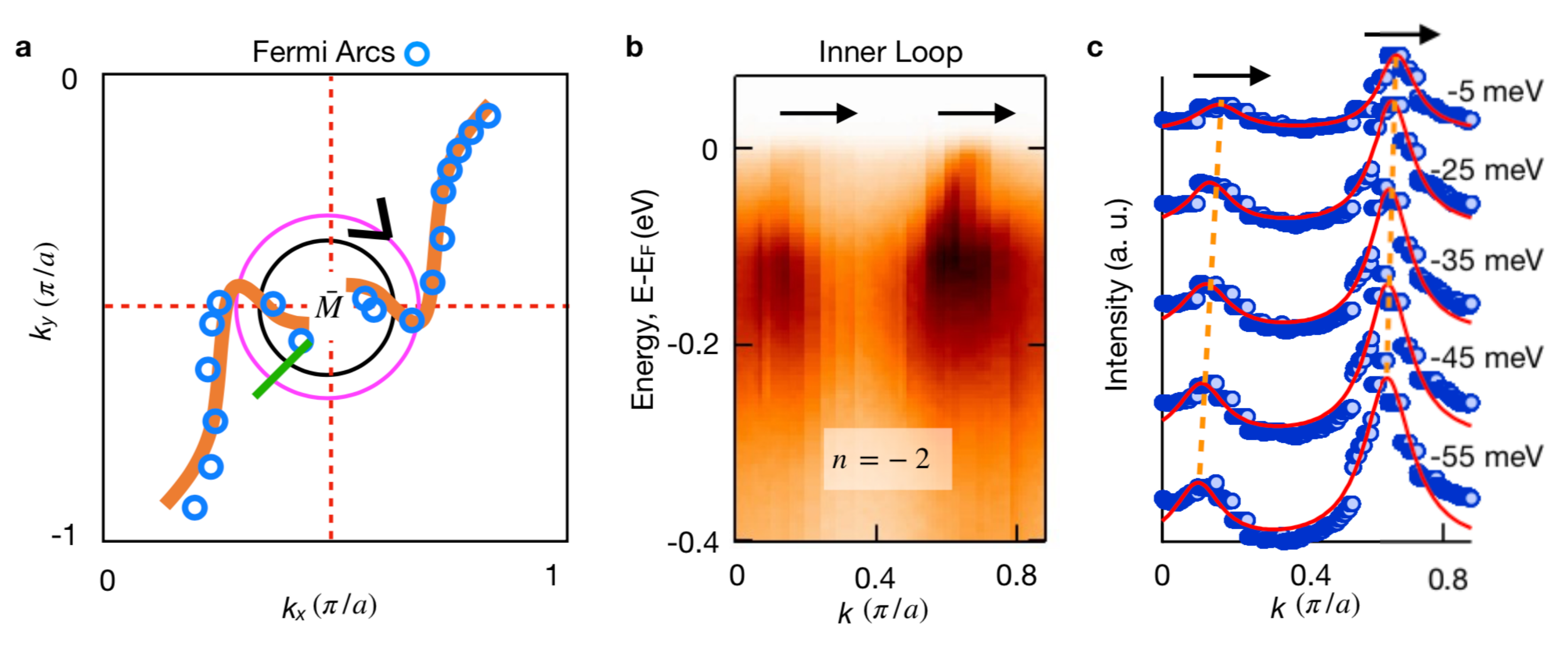}
\caption{\textbf{Systematics for the helicoid fitting in CoSi.} \textbf{a}, Fermi arc trajectory extracted from Lorentzian fits to the MDCs (blue open circles) near $\bar{M}$ with overlaid schematic of the observed features (orange lines). There are two closed contours enclosing $\bar{M}$ on which we count chiral edge modes, the outer loop (magenta, Fig.~\ref{Fig4}) and the inner loop (black). \textbf{b}, Energy-momentum cut along the inner loop, radius $0.14\pi/a$ starting from the green notch in (a) and winding clockwise, black arrow in (a). \textbf{c}, Lorentzian fits (red curves) to the MDCs (blue dots) to extract the helicoid dispersion of the Fermi arcs. We observe two right-moving chiral edge modes dispersing towards \ef, marked schematically (dotted orange lines). The corresponding bulk manifold has Chern number $n=-2$ \cite{Wan}.}
\label{ExtFig_hel_fit}
\end{figure}
\clearpage
\newpage

\begin{figure}
\includegraphics[width=165mm]{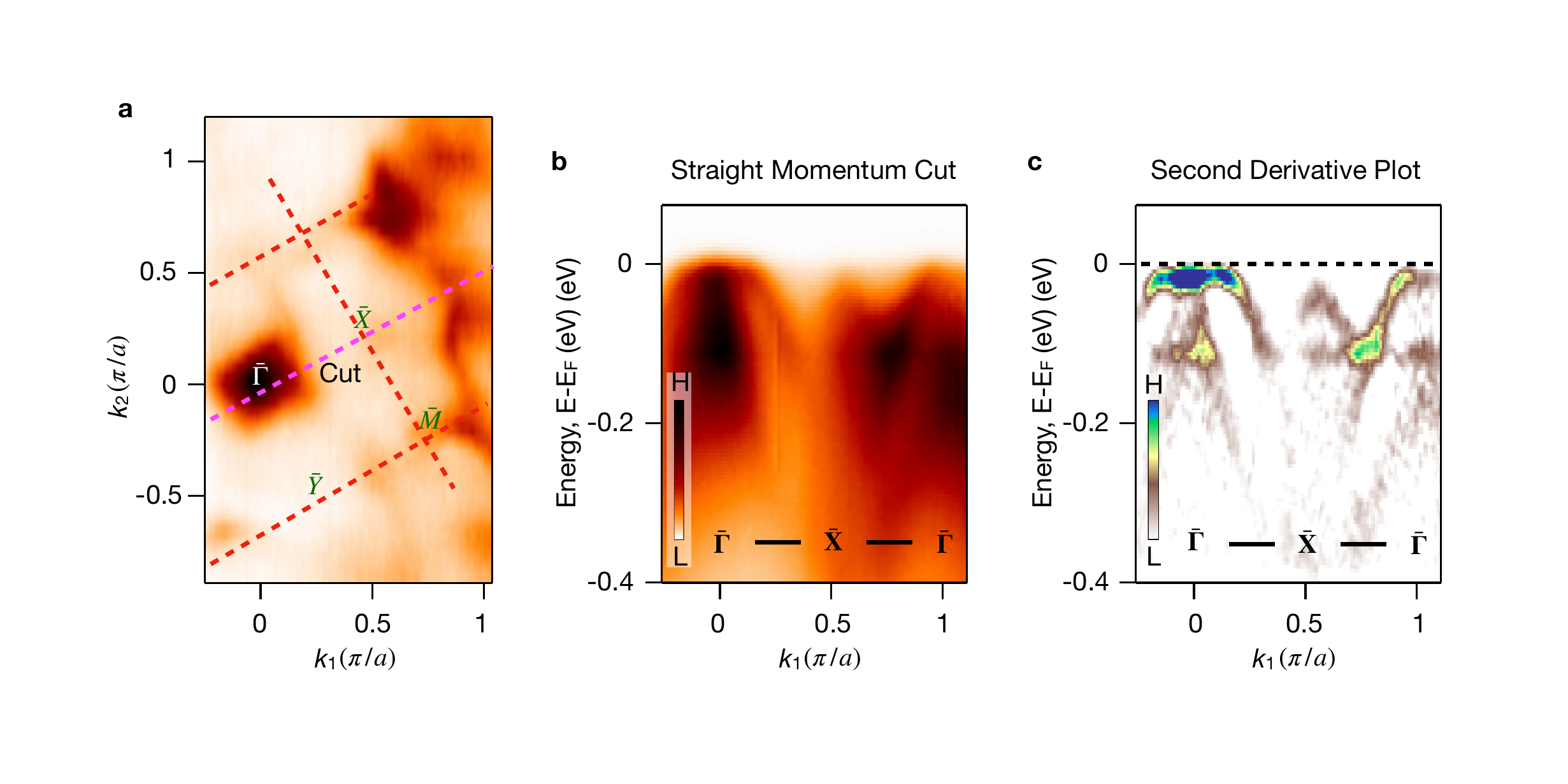}
\caption{\textbf{Surveying the $\bar{\Gamma}$ and $\bar{X}$ pockets on the (001) surface of CoSi.}
\textbf{a}, ARPES Fermi surface with the Brillouin zone boundary (red dotted line) and the location of the energy-momentum cut of interest (magenta dotted line). \textbf{b}, $\bar{\Gamma}-\bar{X}-\bar{\Gamma}$ high-symmetry line energy-momentum cut. \textbf{c}, Second derivative plot of (b). We observe an outer and inner hole-like band at $\bar{\Gamma}$, while at $\bar{X}$ we observe signatures of a single hole-like band.}
\label{ExtFig8_bulk_pockets}
\end{figure}
\clearpage
\newpage

\end{document}